Title: In-orbit Test of the Weak Equivalence Principle with Atom Interferometry

Authors

Dan-Fang Zhang[1,2], Jing-Ting Li[1,2], Wen-Zhang Wang[1,2], Wei-Hao Xu[1,2], Jia-Yi Wei[1,2], Xiao Li[1], Yi-Bo Wang[1], Dong-Feng Gao[1], Jia-Qi Zhong[1,3], Biao Tang[1], Lin Zhou[1,3], Run-Bing Li[1,3], Huan-Yao Sun[1], Qun-Feng Chen[1], Lei Qin[1], Mei-zhen An[4], Zong-Feng Li[4], Shu-Quan Wang[4], Xiao-Xiao Guo[4], Yao Tian[4], Xi-He Yu[4], Hong-En Zhong[4], Xi Chen[1,5*], Jin Wang[1,3,5*], Ming-Sheng Zhan[1,3,5]*

Affiliations

[1]Wuhan Institute of Physics and Mathematics, Innovation Academy for Precision Measurement Science and Technology, Chinese Academy of Sciences; Wuhan 430071, China.

[2]School of Physical Sciences, University of Chinese Academy of Sciences; Beijing 100049, China.

[3]Hefei National Laboratory; Hefei 230088, China.

[4]Technology and Engineering Center for Space Utilization, Chinese Academy of Sciences; Beijing 100094, China.

[5]Wuhan Institute of Quantum Technology; Wuhan 430206, China.

*Corresponding author. Email: chenxi@apm.ac.cn; wangjin@apm.ac.cn; mszhan@apm.ac.cn

Abstract:

The Weak Equivalence Principle (WEP) is a central pillar of general relativity. Its precise test with quantum systems in space offers a unique window onto new physics. Here we report the first in-orbit quantum test of the WEP. A dual-species ($^{85}$Rb/$^{87}$Rb) atom interferometer is realized aboard the China Space Station. Methods of platform motion suppression, fluorescence detection switching, and two-photon detuning switching are developed to eliminate phase noise and improve measurement accuracy. A test uncertainty of $2.8\times10^{-8}$ is obtained from 280 days of WEP test data, and a test result of $(-3.1\pm4.6)\times10^{-7}$ is achieved after error estimation. This improves prior atom-interferometric WEP tests in microgravity by three orders of magnitude. This work paves the way for space-borne quantum inertial sensors and their application to future fundamental physics in space.

Teaser

Weak equivalence principle test using space-based atom interferometer with a precision three orders of magnitude higher than prior microgravity atom interferometry.

MAIN TEXT

Introduction

The Weak Equivalence Principle (WEP), which posits the universality of free fall, is a cornerstone of Einstein's theory of general relativity. Its continued validation through ever more precise experiments is essential, as any detected violation would herald new physics (*1*). Tests using macroscopic bodies have reached phenomenal precision, at the $10^{-13}$ level on Earth (*2,3*) and $10^{-15}$ in space (*4,5*). Owing to its quantum nature and high-precision potential, cold atom interferometry (CAI) has found applications in inertial sensing and fundamental physics research (*6-16*), and has emerged as a powerful technique for testing the WEP (*17-31*). By extending the free-fall time in large towers, CAI has achieved WEP test precision on the order of $10^{-12}$ (*29*). CAI also broadens the scope of these tests by enabling comparisons of atoms in different internal quantum states, such as those differing in spin (*20,24*), superposition (*25*), or hyperfine structure(*28,30*).

The precision of CAI scales quadratically with the interference time *T*. On Earth, the interference time is limited to a few seconds due to gravitational acceleration. Microgravity environments remove this constraint, enabling extended interference times. Pioneering experiments on microgravity platforms, including drop towers (*32,33*), parabolic flight aircraft (*34,23*), sounding rockets (*35,36*), and the Einstein Elevator (*37-39*), have demonstrated the potential of this approach, with the first microgravity CAI-based WEP test reaching a precision of $10^{-4}$ (*23*). However, these platforms offer microgravity durations on the order of seconds or minutes. Reaching an extremely high WEP test precision, e.g., $10^{-17}$ (*40,41*), requires a permanent microgravity environment for long-term data accumulation and comprehensive control of systematic errors. An orbital laboratory, such as the China Space Station (CSS) (*42,43*) or the International Space Station (ISS) (*44-46*), provides the necessary stable environment to realize this potential. Here, we report on the realization of a dual-species atom interferometer onboard the CSS (Fig. 1a) and present the first in-orbit test of the WEP using CAI.

## Results
### Experiment Principle and Procedure

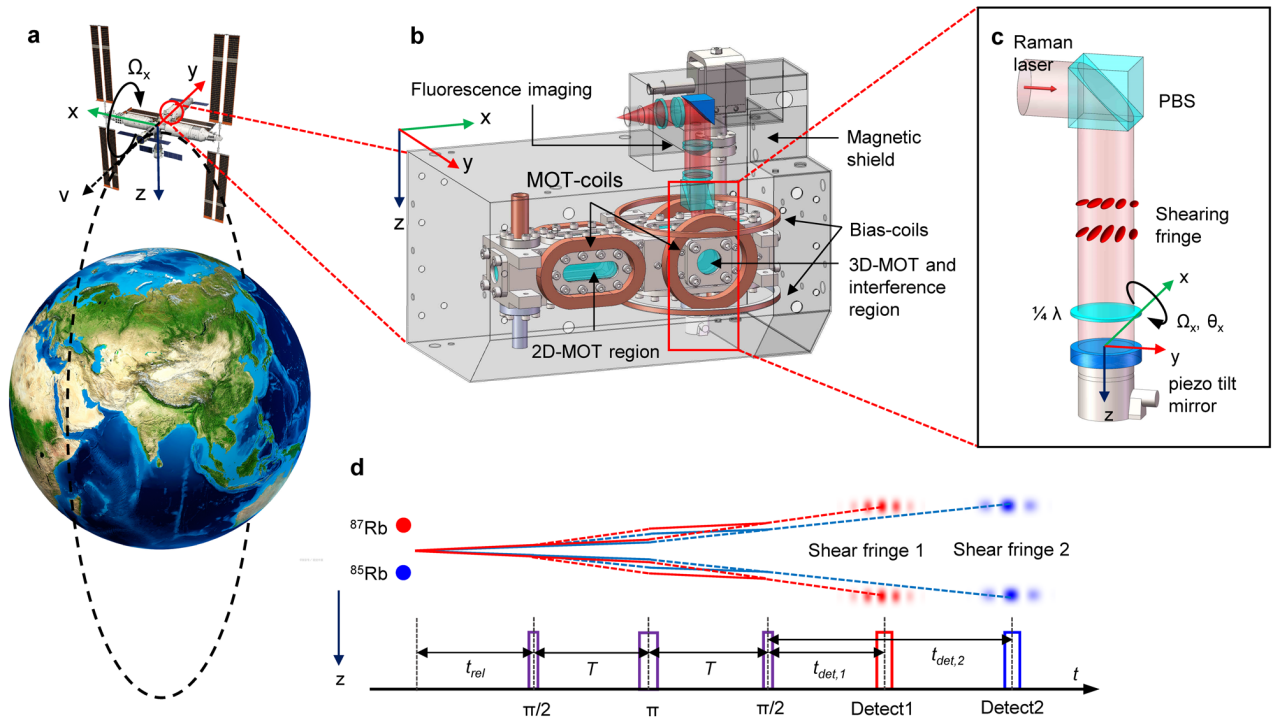

**Fig. 1. Overview of the China Space Station Atom Interferometer (CSSAI). a**, The China Space Station (CSS), its coordinate system, and its orbit around Earth. The CSSAI is installed inside the CSS. **b**, The physical system of the CSSAI and its core functional modules. **c**, Principle of point-source interferometry (PSI) using a piezo-tilt mirror. The shearing fringe is induced by rotation rate of the CSS $\Omega_x$ and also the angle of the piezo tilt mirror $\theta_x$. **d**, Dual-species double single diffraction (DSD) interference sequence for $^{85}$Rb and $^{87}$Rb. The fluorescence of the isotopes is detected in sequence to obtain separate spatial interference fringes.

The China Space Station Atom Interferometer (CSSAI) is a compact, dual-species instrument installed within the CSS's High Microgravity Level Research Rack (HMLR) (*42*). Its core system (Fig. 1b) simultaneously prepares and coherently manipulates clouds of $^{85}$Rb and $^{87}$Rb atoms. The interferometer operates on the principle of double single diffraction (DSD) (Fig. 1d) (*23*), driven by counter-propagating Raman lasers that are reflected by a piezo tilt mirror. This configuration creates two symmetric interference loops per isotope. A critical feature of the setup is the point-source interference (PSI) scheme (Fig. 1c) (*47-49*), which transforms the interferometric phase into a spatial fringe pattern, allowing robust extraction of inertial signals. A more detailed description of the experimental procedure is provided in Materials and Methods I.

For our experiment, the optimized atom number and temperature are 3.4×10$^8$ and 6.6 μK for $^{85}$Rb, and 2.9×10$^8$ and 4.5 μK for $^{87}$Rb. The Raman lasers interact with the cold atom cloud starting at $t_{\text{rel}}$=43 ms after the atom clouds are released. The interference time is $T$=50 ms for both isotopes. The single-photon detunings $\delta_{\text{sp}}$ of the Raman laser are 440 MHz and 425 MHz relative to the $^{85}$Rb $F$=3->$F'$=4 D$_2$ transition and the $^{87}$Rb $F$=2->$F'$=3 D$_2$ transition, respectively. The two-photon detuning is labeled as $\delta_{\text{tp}}$. The width of the π Raman pulse is $\tau$=30 μs. The fluorescence is captured through an imaging system with a magnification of 0.45 and recorded with an sCMOS camera. The fluorescence images of $^{85}$Rb and $^{87}$Rb would overlap if detected simultaneously. Therefore, we excite and detect their fluorescence sequentially at $t_{\text{det},1}$=40 ms and $t_{\text{det},2}$=85 ms relative to the last Raman pulse to obtain separate spatial interference images (Fig. 1d).

By fitting the phases of the spatial fringes, the residual acceleration values along the z-direction are obtained (*43*). The WEP violation coefficient $\eta_{\text{Rb85,Rb87}}$ is defined as:

$$\eta_{\text{Rb85,Rb87}} = \frac{a_{z,\text{Rb85}} - a_{z,\text{Rb87}}}{g_z} = \frac{\Delta\phi_\eta}{\bar{k}_{\text{eff}} g_z T^2} \tag{1}$$

where $a_{z,\text{Rb85}}$ and $a_{z,\text{Rb87}}$ are the residual accelerations measured by the $^{85}$Rb and $^{87}$Rb CAIs, $g_z$ is the earth's gravitational acceleration at the position of the CSS along the z-direction and is 8.72 m/s$^2$ according to the altitude of the CSS. $\eta_{\text{Rb85,Rb87}}$ is derived from the WEP-violation-induced differential phase $\Delta\phi_\eta$ of the interference fringes, where $\bar{k}_{\text{eff}}$ is the averaged Raman laser's wave vector of the two Rb isotopes.

# Platform Motion Effect Suppression

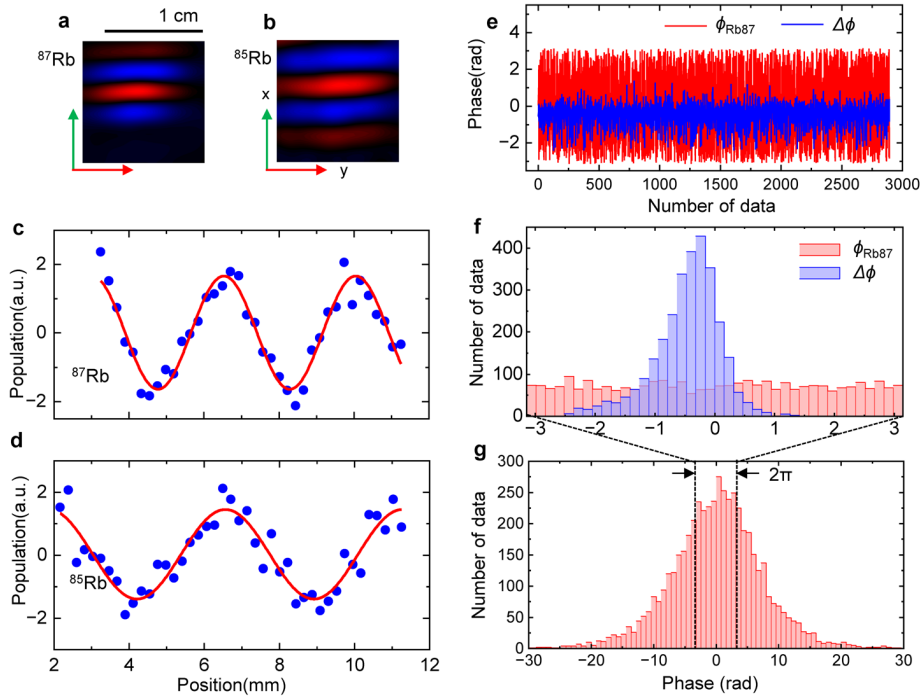

**Fig. 2. Dual-species interference fringes and differential phase extraction. a**, **b**, Principal component analysis (PCA) of the fluorescence images for a, $^{87}$Rb and b, $^{85}$Rb, showing 2D shearing fringes. **c**, **d**, Typical 1D interference fringes for **c**, $^{87}$Rb and **d**, $^{85}$Rb, obtained by averaging the 2D images in the x-direction and using a normalization method (*43*). **e**, Time series of the measured interference phase $\phi_{Rb87}$ and the differential phase $\Delta\phi$. **f**, Histograms of $\phi_{Rb87}$ and $\Delta\phi$. **g**, Histogram of the acceleration induced interference phase calculated according to the acceleration recorded by a classical accelerometer.

The CSS maintains a nadir-pointing attitude with a constant rotation around the x-axis (Fig. 1a). The x-axis rotation $\Omega_x$ has a mean value of −1.138 mrad/s with a standard deviation of 12.0 μrad/s over the long-term measurements, and the rotation rates in the other two axes are nearly zero. By precisely controlling the angles of the piezo tilt mirror during the Raman pulses according to an optimized relation (*29,43*), the decoherent effect induced by the atom cloud's position distribution is eliminated, and proper fringe spatial periods are introduced for interference phase extraction. The optimized angle relation is related to the detection time $t_{det}$, and is different for the two Rb isotopes. We choose an effective detection time between $t_{det,1}$ and $t_{det,2}$ to calculate and set the optimized angles. Clear dual-species interference fringes are shown in Fig. 2a-2d. The spatial frequencies $f_{spa}$ of the observed fringe patterns are 1.72 rad/mm and 1.36 rad/mm for $^{87}$Rb and $^{85}$Rb, respectively. This frequency difference arises from the different fluorescence detection times. Without angle compensation, the rotation of the CSS leads to a fringe frequency of about 0.5 rad/mm, which results in a phase variation less than $2\pi$ in the 1.1cm image width. This makes it difficult to extract the interference fringes. The fringe contrast is also lower and the phase error is higher compared with the rotation-compensated case.

The CSS's platform experiences residual acceleration due to the non-conservative force acting on the station, crew activities, and vibrations from scientific racks. The phases for a single-species atom interference fringes are measured, as shown in Fig. 2e. Their range is restricted to $2\pi$ due to the value range of the inverse sine function. The phase exhibits a uniform distribution, as shown in Fig. 2f, suggesting that its true distribution is much broader than $2\pi$. For comparison, the acceleration of the CSSAI was measured by a classical accelerometer, and the corresponding phase was calculated from the measured acceleration using the sensitivity function (Supplementary Materials). The phase fluctuations for an individual interferometer are shown in Fig. 2g, and have a standard deviation of 7.8 rad. The differential phase $\Delta\phi = \phi_{\text{Rb85}} - \phi_{\text{Rb87}}$ is measured and is shown in Fig. 2e, 2f. It has a standard deviation of 0.55 rad—over an order of magnitude smaller than the single-species phase noise. This is because the dual-species interferometer, operating with an identical time sequence, inherently provides a high vibration rejection ratio on the order of $(k_{\text{eff,Rb87}} - k_{\text{eff,Rb85}})/\bar{k}_{\text{eff}} \sim 10^{-6}$. The residual noise of the differential phase is mainly attributed to the fringe's signal-to-noise ratio rather than the acceleration.

**Suppression of the Spatial Frequency Induced Phase**

The differential phase has a significant offset from zero, as shown in Fig. 2e, 2f. One reason for this offset is the spatial frequency mismatch of the interference fringes. The interference phases of the dual fringes are obtained through sine fitting. An offset of the initial fitting position $y_0$ will cause phase offsets as $f_{\text{spa}}y_0$. When the spatial frequencies are different, these phase offsets are not identical, and a residual differential phase exists. The zero point of $y_0$ is at the position of the piezo tilt mirror's rotation axis and is hard to estimate accurately, as shown in Fig. 3a.

To solve this problem, we designed a fluorescence detection switching method. Two groups of dual-species interference experiments are conducted with alternated fluorescence detection sequence, then we measure the differential phase $\Delta\phi_i$ for each group and calculate their average $\Delta\phi_{\text{avg}}$, where i=1,2 represents different sequences. This method can effectively eliminate the phase shift induced by the spatial frequency mismatch. To verify its effectiveness, we fit the two groups of interference fringe with changing fitting position $y_0$. The calculated $\Delta\phi_i$ and $\Delta\phi_{\text{avg}}$ are shown in Fig. 3b. The differential phase $\Delta\phi_i$ changes by approximately 1.6 rad when $y_0$ varies by 4.3 mm. However, the averaged differential phase $\Delta\phi_{\text{avg}}$ remains constant with a maximum phase change of 0.044 rad. A differential phase suppression ratio of about 36 is achieved.

We also performed a theoretical analysis to determine the magnitude and suppression ratio of the uncertainty associated with this phase offset. The formulas for $\Delta\phi_i$ and $\Delta\phi_{\text{avg}}$ are derived, and the uncertainties of these phases are calculated based on typical values of the experimental parameters. Detailed information is illustrated in the Materials and Methods II section. The majority terms of the phase uncertainty of $\Delta\phi_i$ are induced by the uncertainty of $y_0$ and also the position uncertainty of the cold atom cloud. The calculated differential phase uncertainty for $\Delta\phi_i$ and $\Delta\phi_{\text{avg}}$ are 0.47 rad and $3.7\times10^{-4}$ rad, respectively, corresponding to a suppression factor of more than 1000 by using the fluorescence detection switching method.

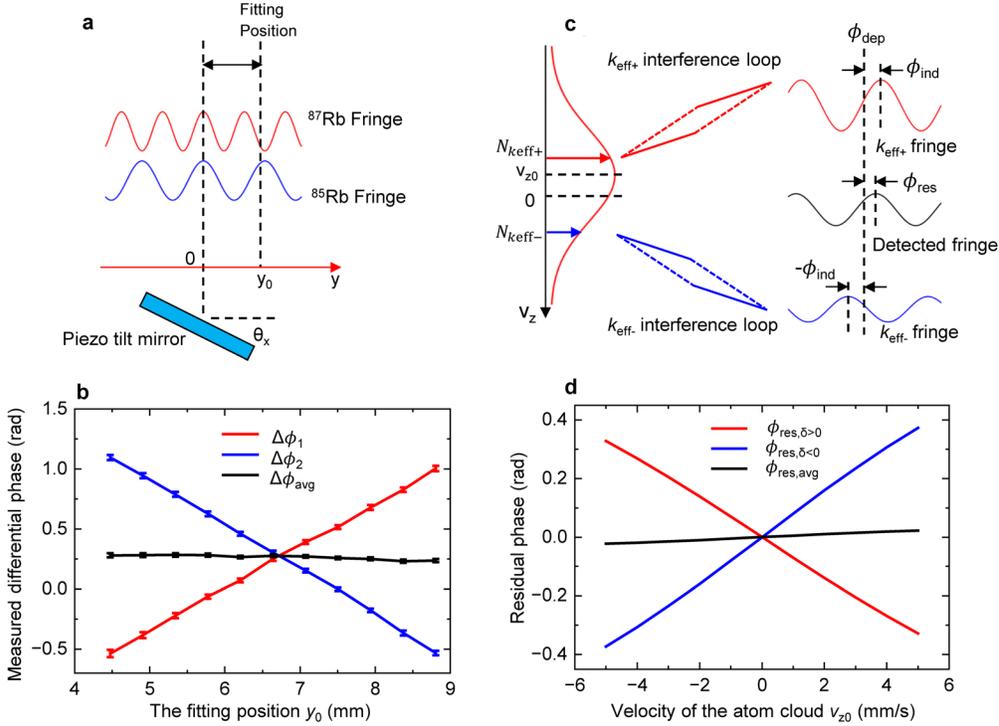

**Fig. 3. Switching methods for fluorescence detection and two-photon detuning. a**, Diagram of the induced differential phase when $f_{spa}$ are different for the dual interferometers. **b**, Calculated differential phase $\Delta\phi_i$ and their averaged $\Delta\phi_{avg}$ with varying sine-fitting offset $y_0$ for the interference fringe. **c**, Diagram of the shearing fringe of $k_{eff\pm}$ and their summation when $v_{z0}$ is not zero. **d**, Calculation of the residual phase $\phi_{res}$ with $\delta_{tp} > 0$ and $\delta_{tp} < 0$ and their average with varying $v_{z0}$. $\phi_{ind}$ is set to 0.92 rad.

## Suppression of the k_eff-Independent Phase

The differential phase is still offset from zero after the fluorescence detection switching method. Another important reason is the asymmetry of the DSD interference loop. The DSD interference scheme employs two interference loops driven by the $k_{eff+}$ and $k_{eff-}$ Raman laser pairs. These loops are identical in shape but opposite in direction. The interference phase of each loop comprises $k_{eff}$-dependent phase $\phi_{dep}$ such as the acceleration related phase, and $k_{eff}$-independent phase $\phi_{ind}$ such as the light shift induced phase. $\phi_{dep}$ shifts the spatial fringes of the two interference loops in the same direction, but $\phi_{ind}$ shift the fringes in opposite directions (*23*). The detected interference fringe is the sum of the fringes of the two loops and has a phase of $\phi_{det}$. If the amplitudes of the two interference loops are the same, $\phi_{ind}$ is eliminated and $\phi_{det}$ is just $\phi_{dep}$. However, if the amplitudes of the two interference loops are different, $\phi_{ind}$ will induce a residual phase shift $\phi_{res}$, as shown in Fig. 3c.

The residual velocity of the atom cloud in the z-direction $v_{z0}$ is the main factor that breaks the symmetry of the interference loops. A non-zero velocity component along the interferometer axis ($v_{z0} \neq 0$) results in different atom numbers ($N_{k_{eff\pm}}$) being selected by the $k_{eff\pm}$ Raman transitions, as illustrated in Fig. 3c. A residual velocity of, e.g., 4 mm/s, would cause a 16% change in the population of the $^{85}$Rb cold atom cloud. The value of $v_{z0}$ cannot be measured directly for our experiment configurations, but its range is estimated to be within ±4 mm/s, as estimated according to the measured velocity of the atom cloud in the x, y directions through the time-of-flight (TOF) method. Detailed illustration is in the Supplementary Materials.

To eliminate $\phi_{\text{res}}$, we designed a two-photon detuning switching method. Two groups of dual-species interference experiments are conducted with alternated sign of the two-photon detuning $\delta_{\text{tp}}$. $N_{k_{\text{eff}\pm}}$ selected by the $k_{\text{eff}\pm}$ transitions are then reversed for these two configurations, and the signs of $\phi_{\text{res}}$ are reversed too. An average of the detected phases $\phi_{\text{det}}$ for these two configurations will eliminate the residual phase $\phi_{\text{res}}$. Numerical simulation of the DSD interference process for the PSI fringe (Materials and Methods III) is developed to calculate the phase error elimination ratio of this method. We set the $k_{\text{eff}}$-independent phase $\phi_{ind}$ to be 0.92 rad, and vary $v_z$ from −5 mm/s to 5 mm/s. Then we calculate $\phi_{\text{res}}$ for $\delta_{\text{tp}} > 0$ and $\delta_{\text{tp}} < 0$ and their average according to the actual experiment parameters of $^{87}$Rb CAI. The results are shown in Fig. 3d. The residual phases for individual configurations exhibit opposite slopes, with a maximum amplitude of 0.65 rad. However, the averaged residual phase has a maximum amplitude of 0.04 rad, which shows a residual phase suppression ratio about 15.

## Long Term Differential Phase Measurement

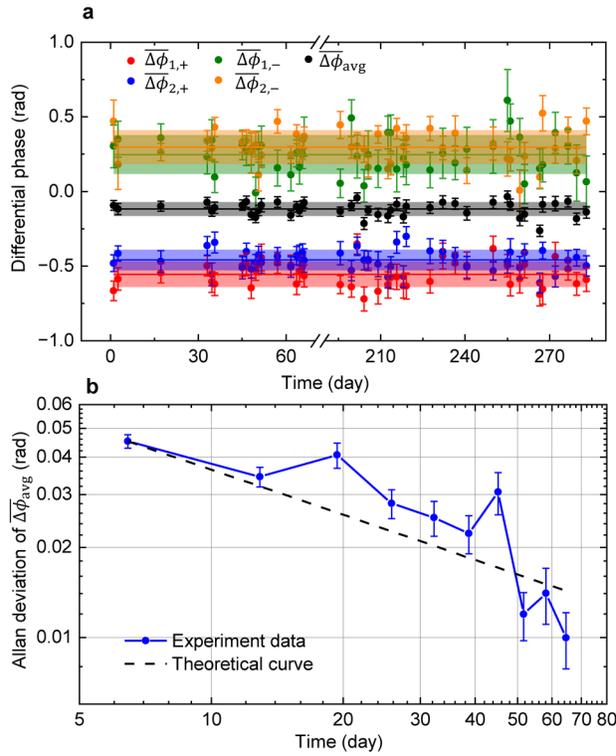

**Fig. 4. Long-term differential phase measurement and stability**. **a**, Averaged differential phase $\overline{\Delta\phi}_{i,j}$ for the four measurement configurations and their average $\overline{\Delta\phi}_{\text{avg}}$ over long-term measurement. The starting date is August 27, 2024. The dots represent the values of experiment data. The solid lines represent the data's average values. The dashed areas represent the data's standard deviation. **b,** Allan deviation of $\overline{\Delta\phi}_{\text{avg}}$ over the experiment time $t$, demonstrating the stability of the differential phase measurement. The dots represent the calculated Allan deviation. The dashed line represents the theoretical curve. It is calculated by dividing the Allan deviation of the first data point by $\sqrt{t/t_{\text{sin}}}$, where $t_{\text{sin}}$ is the effective time interval of a single differential phase.

Using the two switching methods, we conducted a long-term WEP test with four experimental configurations. Over a period of 280 days, we acquired more than 9700 pairs of interference fringes. The differential phases of each configuration are averaged to 44 data points to reduce the noise. The averaged differential phases are denoted as $\overline{\Delta\phi}_{i,j}$, where $i$=1,2 represents different sequences, and $j$=+,− represents different sign of $\delta_{\text{tp}}$. The averaged phase $\overline{\Delta\phi}_{avg} = \sum_{i,j} \overline{\Delta\phi}_{i,j}/4$ are also calculated. The calculated phases are shown in Fig. 4a. The differential phase bias is reduced from a maximum value of −0.556 rad for individual configurations to an average value of −0.117 rad, demonstrating the effectiveness of the switching methods and its critical role in obtaining a reliable WEP test result.

The averaged differential phases $\overline{\Delta\phi}_{avg}$ exhibit a standard deviation of 0.046 rad. The long-term stability of the differential phase is represented by the Allan deviation. It reaches 0.010 rad for the 10 data points averaging, which represents an effective averaging time of 64 days, as shown in Fig. 4b. This indicates that the measured differential phases are stable over months, and reach a WEP test resolution of $2.8\times10^{-8}$ according to Eq. 1.

**Error Estimation**

A detailed error estimation is essential for obtaining an accurate WEP test result. The major systematic error terms are summarized in Table 1. The dominant uncertainty arises from a newly identified effect: imaging angle offset, which results from the coupling between the centroid offset of the DSD interference fringes and the angular misalignment of the imaging system. The centrifugal force and gravity gradient effects are both linked to the centroid offset of the DSD interference loops, which is primarily induced by the residual velocity $v_{z0}$. The single-photon ac Stark shift introduces a $k_{\text{eff}}$-independent phase and is effectively suppressed by the two-photon detuning switching method. Phase shifts induced by residual acceleration, the multiple-sideband effect of the Raman lasers, and uncertainties in $k_{\text{eff}}$ and $T$ are all proportional to the local gravitational acceleration, which are suppressed by at least three orders of magnitude compared to ground-based measurements. A detailed discussion of each systematic effect is provided in the Materials and Methods IV. The measured average differential phase $\overline{\Delta\phi}_{avg}$ after error correction is (−0.11±0.16) rad. The WEP violation coefficient $\eta_{\text{Rb85,Rb87}}$ is then calculated to be $(-3.1\pm4.6)\times10^{-7}$ according to Eq. 1.

Table 1. Error budget for the differential phase.

| Terms | | Differential phase(rad) |
|---|---|---|
| Measurement result | | −0.12±0.01 |
| Systematic effects | Residual acceleration | $(-2.1\pm8.0)\times10^{-5}$ |
| | Centrifugal force | $(0.0\pm7.4)\times10^{-5}$ |
| | Gravity gradient | $(0.0\pm1.5)\times10^{-4}$ |
| | Imaging angle offset | 0.01±0.16 |
| | Shearing fringe frequency mismatch | $(-1.5\pm3.7)\times10^{-4}$ |
| | Single-photon ac Stark shift | $(0.0\pm1.0)\times10^{-2}$ |
| | Two-photon ac Stark shift | $(0.0\pm2.3)\times10^{-3}$ |
| | Multiple-sideband effect | $(0.0\pm2.2)\times10^{-5}$ |
| | Magnetic field | $(0.0\pm3.8)\times10^{-4}$ |
| | Wavefront distortion | $(0.0\pm1.2)\times10^{-2}$ |
| | Uncertainty of $k_{\text{eff}}$ and $T$ | $<10^{-6}$ |
| EP violation induced phase | | −0.11±0.16 |

**Discussion**

This paper presents the WEP test experiment conducted with the CSSAI. This work successfully transitions such experiments from short-duration microgravity platforms to a permanent orbital laboratory. The instrument operated stably for over 280 days. We addressed the challenges posed by the significant rotation rate and residual acceleration of the CSS platform. Methods are designed to suppress the differential phase errors. Error estimations are carried out to give a test result with a precision of $4.6\times10^{-7}$, improving upon the best previous microgravity test result based on atom interferometry by three orders of magnitude.

This work not only demonstrates the feasibility of conducting high-precision scientific experiments with cold atom interferometers in space, but also serves as a valuable reference for future space-based WEP tests aiming for even higher precision(*41,50,51*). By rotating the spacecraft with a certain spin velocity, one can modulate the WEP signal and suppress the noise at the orbital frequency (*4*). For this case, the proposed PSI interference scheme with optimized angle relation could be used to extract the interference phase and reduce the atom parameter induced phase error. The spacecraft could reduce the residual acceleration by the drag-free technology (*4,52*). By applying scale-factor matching or Lissajous curve fitting to the dual interferometer (*41*), the residual acceleration-induced WEP test error could be further suppressed. The proposed two-photon detuning switching method eliminates the $k_{\text{eff}}$-independent phase and can improve the measurement accuracy for the Raman or single-photon diffraction interference scheme (*53*). This study also provides a comprehensive error analysis, identifying several newly discovered sources of phase error and highlighting differences between ground-based and space-based experiments. This provides a reference for future space-based cold atom interferometer design and error analysis. As a next step, deeply cooled atom ensembles, such as delta-kick-cooled ultracold quantum gases, could greatly extend the interference time and thus improve the WEP test resolution (*32,35,44*).

This work reports the first WEP test using an atom interferometer in orbit, validating key technologies for quantum precision measurement in space. It marks the transition of space-based cold atom interferometry from proof-of-principle demonstrations to practical applications, opening new avenues for fundamental physics research in space.

## Materials and Methods
## I Experimental Design

The CSSAI operates in an intermittent mode on the Free-floating Platform for Microgravity Experiment (FPME) within the HMLR (*43*). For a single round of WEP test experiments, the total operation time is limited to 70 minutes to prevent overheating of the payload. The first 13 minutes are dedicated to the startup process, which includes stabilization of temperature control, achievement of a stable vacuum level, and automatic locking of the laser frequencies. The subsequent 18 minutes are allocated to scientific experiments, during which the atom interference sequence is executed periodically. WEP test experiments with four experimental configurations are performed alternately, and approximately 30 pairs of interference fringes are obtained per configuration per round. The final 39 minutes are reserved for data transfer.

## II Analysis of the Phase for the Shearing Fringe
## (1) Phase of the PSI Interference Fringe

The formula for shearing fringe when considering the acceleration, rotation and Raman mirror's angle is (*43*)

$$\phi = k_{\text{eff}} a_z T^2 + k_{\text{eff}}\left[2\Omega_x v_y T^2 + \theta_{x,1} r_y + \theta_{x,3}(r_y + 2v_y T)\right], \qquad (2)$$

where $k_{\text{eff}}$ is the effective wave vector, $T$ is the time interval between the Raman pulses, $a_z$ is the residual acceleration along the z-axis, $\Omega_x$ is the rotation rate of the CSS around the x-axis, and $\theta_{x,1}$ and $\theta_{x,3}$ are the angles of the Raman laser beam along the x-axis at the time of the 1st and 3rd Raman laser pulses, relative to its angle at the time of the 2nd Raman laser pulse. $r_y$ and $v_y$ are the position and velocity of the atom at the time of the 1st Raman laser pulse.

When $\theta_{x,1}$ and $\theta_{x,3}$ have the following relationship (43),
$$\theta_{x,1} = \theta_{xo,1} = \frac{-t_{det}\theta_{x,3} + 2\Omega_x T^2}{2T + t_{det}}, \tag{3}$$
the decoherent effect induced by the atom cloud's distribution is eliminated, where $t_{det}$ is the time delay between the last Raman laser pulse and the detection laser pulse. The shearing fringe has the following form (43)
$$\phi = f_{yo}R_y, \tag{4a}$$
$$f_{yo} = \frac{2k_{eff}}{2T + t_{det}}(\theta_{x,3}T + \Omega_x T^2), \tag{4b}$$
where $R_y = r_y - v_y(2T + t_{det})$ is the position of the atom at the detection time. $f_{yo}$ is the optimized spatial frequency. When the relation of Eq. 3 is not exactly fulfilled and has an angle offset $\Delta\theta_x = \theta_{x,1} - \theta_{xo,1}$, then both the phase and the spatial frequency have to be modified.
$$\phi = f_y R_y + k_{eff}[\alpha(\beta\rho_{y0} - v_{y0}t_{tot})\Delta\theta_x, \tag{5a}$$
$$f_y = f_{yo} + k_{eff}\left(\frac{t_{rel}}{t_{tot}} + \alpha\right)\Delta\theta_x. \tag{5b}$$
where $\alpha = \frac{t_{tot} - t_{rel}}{t_{tot}} \cdot \frac{\sigma_{\rho y}^2}{\sigma_{vy}^2 t_{tot}^2 + \sigma_{\rho y}^2}$ and $\beta = \sigma_{vy}^2 t_{tot}^2 / \sigma_{\rho y}^2$, $t_{rel}$ is the time from atom release to the first Raman pulse, $t_{tot} = t_{rel} + 2T + t_{det}$ is the total time. $\sigma_{\rho y}, \sigma_{vy}$ are the distribution widths of the position and velocity of the atom cloud. $\rho_{y0}$ and $v_{y0}$ are the initial position and initial velocity of the atom cloud.

## (2) Differential Phase of the PSI Interference Fringe

For the dual-species atom interferometer WEP test experiment, the fluorescence detection times for the two isotopes are different, denoted as $t_{det,1}$ and $t_{det,2}$. This will lead to different ideal angles $\theta_{xo,1}$ according to Eq. 3. Therefore, we use a time $t_{det}$ midway between $t_{det,1}$ and $t_{det,2}$ to set the angle of $\theta_{x,1}$ according to Eq. 3. Then the modified phase of Eq. 5 is denoted as $\phi(i,j)$, where $i=1,2$ for $^{85}$Rb and $^{87}$Rb isotopes and $j=1,2$ for detection times $t_{det,1}$ and $t_{det,2}$. The differential phase $\Delta\phi_1$ for the case where $^{85}$Rb is detected before $^{87}$Rb, and $\Delta\phi_2$ for the case where $^{87}$Rb is detected before $^{85}$Rb are
$$\Delta\phi_1 = \phi(1,1) - \phi(2,2), \tag{6a}$$
$$\Delta\phi_2 = \phi(1,2) - \phi(2,1). \tag{6b}$$
After derivation, $\Delta\phi_1$ has the form
$$\Delta\phi_1 = k_{eff}[\alpha_{1,1}(\beta_{1,1}\rho_{1,y0} - v_{1,y0}t_{tot,1})\Delta\theta_{x,1} - \alpha_{2,2}(\beta_{2,2}\rho_{2,y0} - v_{2,y0}t_{tot,2})\Delta\theta_{x,2}]$$
$$+ k_{eff}\Delta f_y y_0 + k_{eff}[(t_{rel}/t_{tot,1} + \alpha_{1,1})\Delta\theta_{x,1} - (t_{rel}/t_{tot,2} + \alpha_{2,2})\Delta\theta_{x,2}]y_0, \tag{7}$$
where, $\alpha_{i,j}, \beta_{i,j}, \rho_{i,y0}, v_{i,y0}, t_{tot,j}, \Delta\theta_{x,j}$ are defined parameters with $i=1,2$ for $^{85}$Rb and $^{87}$Rb isotopes and $j=1,2$ for detection times $t_{det,1}$ and $t_{det,2}$. $\Delta f$ is the differential spatial frequency with $\Delta f_y = f_{yo,1} - f_{yo,2}$. The zero point of $R_y$ is at the position of the rotation axis of the tilt mirror, and $y_0$ is the fitting position of the shearing fringe from this zero point, as shown in Fig. 3a. $\Delta\phi_2$ has similar form.

The average differential phase $\Delta\phi_{avg}$ for $\Delta\phi_1$ and $\Delta\phi_2$ are
$$\Delta\phi_{avg} = (\Delta\phi_1 + \Delta\phi_2)/2 \tag{8}$$
After derivation, $\Delta\phi_{avg}$ has the form
$$\Delta\phi_{avg} = \frac{1}{2}k_{eff}\{[(\alpha_{1,1}(\beta_{1,1}\rho_{1,y0} - v_{1,y0}t_{tot,1}) - \alpha_{2,1}(\beta_{2,1}\rho_{2,y0} - v_{2,y0}t_{tot,1}))\Delta\theta_{x,1}$$
$$+ (\alpha_{1,2}(\beta_{1,2}\rho_{1,y0} - v_{1,y0}t_{tot,2}) - \alpha_{2,2}(\beta_{2,2}\rho_{2,y0} - v_{2,y0}t_{tot,2}))\Delta\theta_{x,2}]$$
$$+ [(\alpha_{1,1} - \alpha_{2,1})\Delta\theta_{x,1} + (\alpha_{1,2} - \alpha_{2,2})\Delta\theta_{x,2}]y_0\}. \tag{9}$$

Two features can be seen for $\Delta\phi_{avg}$. First, the phase term $k_{eff}\Delta f_y y_0$ is eliminated due to the fluorescence detection switching. Second, if the two atom clouds have identical distributions, that is $\alpha_{1,j} = \alpha_{2,j}$, $\beta_{1,j} = \beta_{2,j}$, $\rho_{1,y0} = \rho_{2,y0}$, $v_{1,y0} = v_{2,y0}$, then the phase error of $\Delta\phi_{avg}$ is completely suppressed according to Eq. 9. If the distributions of the atom clouds are not the same, residual phase error exists.

### (3) Error Estimation for the Differential Phase

Actual experimental parameters are used to calculate the differential phase errors. The value of $k_{eff}$ is calculated through the Raman laser's single photon detuning. The distributions of atom clouds are calculated through the fluorescence images and the time-of-flight (TOF) measurement. The angle of piezo tilt mirror is calculated through its control voltages and the calibrated angle coefficient (43). The timing sequence is determined by its set values. The uncertainty of position of the atom cloud $\rho_{y0}$ and the fitting position offset $y_0$ are restricted by the uncertainty of it coordinate origin, which is the position of the rotation axis of the tilt mirror. By substituting these parameters into Eq. 7 and 9, the differential phases are calculated and are shown in Table 2. The calculated value for $\Delta\phi_1$ is (0.00±0.47) rad. The majority uncertainties come from the $\rho_{y0}$ and $y_0$, where the spatial frequency difference couples to the position uncertainty and contributes phase uncertainty. The calculated value for $\Delta\phi_{avg}$ is $(-1.5\pm3.7)\times10^{-4}$ rad. By applying the fluorescence detection switching method, the uncertainty of the differential phase is suppressed by over 1000 times.

**Table 2.** Error estimation of the differential phase $\Delta\phi_1$ and $\Delta\phi_{avg}$ based on actual experimental parameters.

| Error Source | | Parameters | $\Delta\phi_1$ (rad) | $\Delta\phi_{avg}$ (rad) |
|---|---|---|---|---|
| Phase Offset | | | $1.2\times10^{-3}$ | $-1.5\times10^{-4}$ |
| Phase Offset Uncertainty | Raman wave vector (/m) | $k_{eff}$=1.61058×10$^7$±0.04 | $2.9\times10^{-12}$ | $3.8\times10^{-13}$ |
| | Time sequence (μs) | $t_{rel}$=43.243±0.01<br>$T$=50.128±0.01<br>$t_{det1}$=40.043±0.01<br>$t_{det2}$=85.043±0.01 | $4.0\times10^{-7}$ | $3.1\times10^{-7}$ |
| | Position of the atom cloud (mm) | $\rho_{1,y0}=\rho_{2,y0}$=0.0±1.0 | 0.34 | $2.4\times10^{-4}$ |
| | Distribution of atom cloud | $\sigma_{1,\rho y}=\sigma_{2,\rho y}$=(0.442±0.004) mm<br>$\sigma_{1,vy}$=(21.8±0.7) mm/s<br>$\sigma_{2,vy}$=(17.7±1.6) mm/s<br>$v_{1,y0}$=(0.6±0.5) mm/s<br>$v_{2,y0}$=(2.1±0.9) mm/s | $1.6\times10^{-3}$ | $1.4\times10^{-4}$ |
| | Angle of piezo tilt mirror (μrad) | $\theta_{x,1}$=-116.1±0.4<br>$\theta_{x,3}$=211.8±0.7 | $3.2\times10^{-5}$ | $3.2\times10^{-5}$ |
| | Rotation (mrad/s) | $\Omega_x$=-1.142±0.012 | $1.7\times10^{-5}$ | $2.5\times10^{-5}$ |
| | Fitting position offset (mm) | $y_0$=0.0±1.0 | 0.34 | $2.4\times10^{-4}$ |
| In total | | | 0.00±0.47 | $(-1.5\pm3.7)\times10^{-4}$ |

# III Simulation for the PSI Fringe

To calculate phase shifts induced by effects that are difficult to analyze theoretically, we developed a simulation program for the interference process. This program starts from the initial state of the atom with velocity distribution, solves the state evolution using the Schrödinger equation with spatial coupled phase such as the rotation and tilted angles induced phase, and computes the shearing interference fringes in the position space. Based on the obtained fringes, the phase shifts induced by various parameters can be calculated. Since the interference processes for the two atomic species are similar, we introduce the procedure of the simulation for one species.

## (1) State Evolution Equation and Hamiltonian

The configuration of the Raman lasers for atom interferometry is shown in Fig.5a. $I_1$, $I_2$ are the incident Raman lasers. They are linearly polarized with the same polarization direction. $I_1$, $I_2$ are reflected with the piezo tilt mirror. A quarter-wave plate before the mirror adjusts the Raman laser polarization so that the reflected and incident beams form pairs with perpendicular linear polarizations. The reflected lasers are labeled as $I_{1,r}$, $I_{2,r}$. $I_1$ and $I_{2,r}$ form the $k_{\text{eff}+}$ Raman laser pair, and $I_2$ and $I_{1,r}$ form the $k_{\text{eff}-}$ Raman laser pair.

The atom is modeled with three internal energy levels, labeled as $|a\rangle$, $|b\rangle$ and $|c\rangle$, where $|a\rangle$, $|b\rangle$ are the two ground states and $|c\rangle$ is the excited state. The Raman laser's two-photon detuning is defined as $\delta_{\text{tp}} = f_1 - f_2 - f_{a,b}$, where $f_1$ and $f_2$ are the frequency of the Raman lasers and $f_{a,b}$ is the frequency difference between state $|a\rangle$ and $|b\rangle$. In the general case, the atom has a momentum p. The $k_{\text{eff}\pm}$ Raman laser drives the atom to a series of momentum states labeled $|a, n\hbar k_{\text{eff}} + p\rangle$ and $|b, n\hbar k_{\text{eff}} + p\rangle$, where n=0,±1… and $p = mv$, as shown in Fig. 5b. For convenience, we denote them as $|a, p, n\rangle$ and $|b, p, n\rangle$. The state amplitudes of the above states are denoted as $c(a, p, n)$ and $c(b, p, n)$.

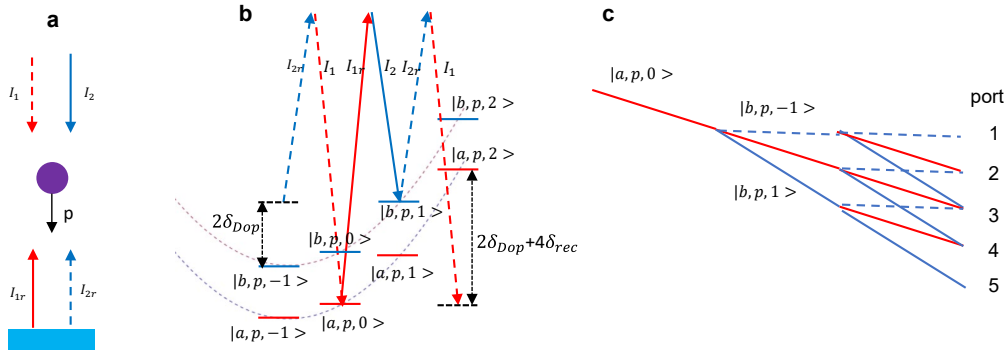

**Fig. 5. Raman lasers and interferometer dynamics for the simulation model. a**, The counter-propagating Raman laser pairs. **b**, Momentum states coupled by the DSD Raman transitions. **c**, Interference loops involving the $|a, p, 0\rangle$ and $|b, p, \pm 1\rangle$ states.

Following the derivation in Ref. (54), the state amplitudes can be calculated by the coupled equations

$$i\hbar \dot{c}(b,p,n) = \frac{\hbar \Omega_{\text{eff}+}}{2} \exp^{i[(2n-1)\delta_{\text{rec}} - \delta_{\text{Dop}} - \delta_{\text{tp}}]t - i\phi_{\text{eff}+}} c(a, p, n-1) + \frac{\hbar \Omega_{\text{eff}-}}{2} \exp^{i[(-2n-1)\delta_{\text{rec}} + \delta_{\text{Dop}} - \delta_{\text{tp}}]t - i\phi_{\text{eff}-}} c(a, p, n+1), \qquad (10a)$$

$$i\hbar\dot{c}(a,p,n) = \frac{\hbar\Omega_{\text{eff}-}}{2}exp^{-i[(-2n+1)\delta_{\text{rec}}+\delta_{\text{Dop}}-\delta_{\text{tp}}]t+i\phi_{\text{eff}-}}c(b,p,n-1) + \\ \frac{\hbar\Omega_{\text{eff}+}}{2}exp^{-i[(2n+1)\delta_{\text{rec}}-\delta_{\text{Dop}}-\delta_{\text{tp}}]t+i\phi_{\text{eff}+}}c(b,p,n+1), \quad (10b)$$

where, $\Omega_{\text{eff}\pm}$ are the effective Rabi frequency of the $k_{\text{eff}\pm}$ Raman transition, $\phi_{\text{eff}\pm}$ are the Raman laser's phases, $\delta_{\text{rec}}$ is the recoil frequency, $\delta_{\text{Dop}} = k_{\text{eff}}v$ is the Doppler frequency shift. Eqs. 10 could be expressed as

$$i\hbar\dot{\varphi}(t) = H\varphi(t), \quad (11)$$

where $\varphi(t) = \sum_n c(a,p,n)|a,p,n\rangle + c(b,p,n)|b,p,n\rangle$.

## (2) Calculation of the PSI Interference Fringe

The Mach-Zehnder interferometer sequence consists of π/2 - π - π/2 Raman pulses. The end state through these pulses is calculated using

$$\varphi_{\text{end}}(t) = U_{\pi/2}U_\pi U_{\pi/2}\varphi_{\text{int}}(t), \quad (12)$$

where $U = e^{-iH/\hbar}$ is the evolution operator.

Equation 12 describes the internal state evolution. However, it does not consider the decoherence effect between different momentum states. Figure 5c shows the interference loops when considering the $|a,p,0\rangle$ and $|b,p,\pm 1\rangle$ states. There are 5 exit ports for the interference loop. If Eq. 12 alone is used to calculate the final state, coherence is assumed between states at different exit ports, as shown in Fig. 5c. However, the finite temperature of the atom cloud causes decoherence between states from different ports. To address this issue, we use a state-grouping method to calculate the population for each exit port individually (*54*). We calculate the state populations for each port, and then calculate the total population over all ports.

To simulate the PSI spatial fringe, the velocity distribution of the atom cloud in the *y* and *z* directions is considered. The velocity distribution is discretized into individual velocity component. For each discrete velocity component in the initial state, Eq. 12 associated with the grouping method is used to calculate the population of the end state. The shearing phases described in Eq. 2 is added to the Raman laser's phase during the calculation. The 2D PSI interference fringes are a sum of the end state population over all velocity components. Typical fringes in the *y-z* plane are shown in Figs. 6a and 6b. A pair of spatial fringes in the y direction and separated along the *z* direction are clearly shown.

To extract the fringe phase, the 2D population is averaged along the z-direction to obtain the 1D shearing fringe, as shown in Figs. 6c and 6d. By fitting these curves with a model of Gaussian envelope modulated by sinusoidal fringe, we can calculate the phase and also other parameters such as the fringe contrast and spatial frequency. Phase shifts induced by various parameters, such as AC Stark shift, atom cloud velocity, rotation, etc., can be calculated.

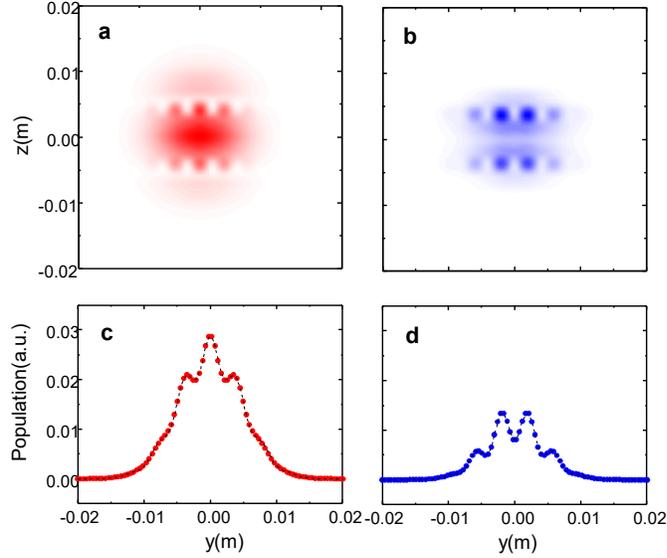

**Fig. 6. Simulation result of the point-source interferometer (PSI) fringes. a, b**, Simulated 2D interference fringes in the y-z plane for the end states $|a\rangle$ and $|b\rangle$. **c, d**, The corresponding 1D spatial fringes obtained by integrating the 2D population distributions along the z-direction. The simulation parameters are: atom cloud temperature 6.7 μK, $t_{rel}$=43 ms, $T$=50 ms, $t_{det}$=40 ms, $\tau$=30 μs, $\delta_{tp}$=69 kHz. $\Omega_x$=-0.0011 rad/s, $\theta_{x,1}$=-93 μrad, $\theta_{x,3}$=188 μrad. The initial velocity of atom cloud $v_{z0}$ is set to 4 mm/s.

## IV Error Estimation
### (1) Acceleration

The differential phase arising from residual acceleration is given by $\Delta\phi = \Delta k_{eff} a_z T^2$, where $\Delta k_{eff}$ is the differential effective wavevector and $a_z$ denotes the residual acceleration measured by a classical accelerometer. $\Delta k_{eff}$ is 6 orders lower than $k_{eff}$, which means the WEP test precision induced by the residual acceleration is also suppressed by the same order. The acceleration is measured by a classical accelerometer. The measured residual acceleration $a_z$ is $(2.7\pm0.6) \times 10^{-4}$ m/s² during the WEP test experiment, and the accelerometer has a zero bias of $\pm 10^{-3}$ m/s². The synthetic residual acceleration value is $(2.7\pm10.0) \times 10^{-4}$ m/s², and its induced differential phase is $(-2.1\pm8.0)\times10^{-5}$ rad.

### (2) Rotation and Piezo Tilt Mirror

Rotation and the tilt angle of the piezo tilt mirror can induce a residual differential phase, even after applying the detection switching method. The derived formula for their influence is illustrated in the Materials and Methods I. By substituting actual experimental parameters, the calculated contribution of $\overline{\Delta\phi}_{avg}$ for this effect is $(-1.5\pm3.7)\times10^{-4}$ rad. The differential phase uncertainty is mainly caused by the uncertainties of the position of the atom clouds and the fit position $y_0$ relative to the position of the tilt mirror's rotation axis. The rotation will also cause centrifugal force. The induced differential phase is $\Delta\phi = -k_{eff}\omega_x^2 \Delta z_{int} T^2$, where $\Delta z_{int} = z_{int,Rb85} - z_{int,Rb87}$ is the centroid position difference of the $^{85}$Rb and $^{87}$Rb interference loops. This centroid position offset is mainly caused by the $v_{z0}$ induced interference loop asymmetry. If $v_{z0}$ is 4 mm/s, the calculated $z_{int}$ is 0.7 mm. Considering the uncertainty of $v_{z0}$ of the two cold atom clouds, the induced differential phase is $(0.0\pm7.4) \times 10^{-5}$ rad.

### (3) Single-Photon Light Shift

The CSSAI uses electro-optic phase modulation to generate laser sidebands that serve as the Raman lasers (*55*). The ratios of the sidebands are monitored by an F-P cavity. According to the measured sideband ratios, the total power of the Raman lasers, and the single-photon detuning of the Raman lasers, the single-photon light shifts $\delta_{spls}$ are calculated to be −27.7 kHz and 38.4 kHz for $^{85}$Rb and $^{87}$Rb. These frequency shifts were verified by in-orbit Raman transition spectroscopy measurements.

$\delta_{spls}$ induced phase is $k_{eff}$-independent for each interference loop of DSD. It causes residual phase $\phi_{res}$ only when the interference loops are asymmetric. The residual phase is calculated to be proportional to $\delta_{spls}$ and $v_{z0}$. Because $\delta_{spls}$ of the two isotopes have opposite signs, the contribution of $\overline{\Delta\phi}_{avg}$ induced by this effect will be maximum when $v_{z0}$ of the two atom clouds have the same sign. By applying the two-photon detuning switching method, the induced differential phase could be suppressed to some extent.

$\delta_{spls}$ induced phase is related to the Raman laser's profile and also the expansion of the cold atom clouds, and it will imprint phase aberration to the 2D-shearing fringe. It is difficult to obtain the analytical derivation. We carried out simulation calculation to obtain the shearing fringe and evaluate the resulted phase shift. $v_{z0}$ of the two atom clouds are both set to 4 mm/s. For the cases of positive and negative two-photon detuning, the calculated differential phases are 0.057 rad and −0.076 rad for the two cases. The induced averaged differential phase is −0.01 rad. Considering the uncertainty of the sign of $v_{z0}$, the induced phase for $\overline{\Delta\phi}_{avg}$ is $(0.0\pm1.0)\times10^{-2}$ rad.

### (4) Two-Photon Light Shift

The two-photon light shift $\delta_{tpls}$ arises from off-resonant Raman transitions. For the DSD interference process, the two-photon detuning is non-zero. When the $k_{eff+}$ Raman transition is in resonance to an atom ensemble with a certain velocity distribution, the $k_{eff-}$ Raman transition is off-resonance to this atom ensemble, and vice versa. $\delta_{tpls}$ is calculated through the effective Rabi frequency and the two-photon detuning (*56*). At the free fall condition on ground, $\delta_{tpls}$ induced phase is mainly induced by the gravity induced Doppler frequency shift, and is a kind of $k_{eff}$-dependent phase (*56*). However, under microgravity condition, the gravity induced Doppler frequency changing is absent, $\delta_{tpls}$ induced phase is found to be $k_{eff}$-independent. Similar to the single-photon light shift case, the induced residual phase exists only when the DSD interference loop is asymmetric. We set $v_{z0}$ for both atom clouds to have a magnitude of 4 mm/s but with opposite sign to maximize the induced differential phase. For the cases of positive and negative two-photon detuning, the calculated differential phases are −2.0 mrad and −2.6 mrad. Considering the uncertainty of the sign of $v_{z0}$, the induced differential phase is $(0.0\pm2.3)\times10^{-3}$ rad.

### (5) Magnetic Field

For the DSD interference, the magnetic field induced phase shift is $\phi = 2\pi\,\alpha B_0 \gamma \hbar k_{eff} T^2 / m_{Rb}$, where $\alpha$ is the second order Zeeman frequency shift coefficient, $B_0$=504 mGal is bias magnetic field, $\gamma$ is its gradient, and $m_{Rb}$ is the mass of rubidium. $\gamma$ is evaluated on ground and has an uncertainty of $(0.0\pm5.6)$ mG/m (*43*). This phase shift is $k_{eff}$-dependent. The induced differential phase is $(0.0\pm3.8)\times10^{-4}$ rad.

**(6) Imaging Angle Offset**

The DSD interference forms a pair of shearing fringes separated in the z-direction. If the populations of the two fringes differ, then the centroid of the fringe is not coincident with its geometric center in the z direction. For a value of $v_{z0}=$ 4 mm/s, the calculated centroid offset $z_{\text{fri}}$ is approximately 1.6 mm. If the imaging system has an angle offset $\theta_z$ in the z direction, it will induce a phase offset of $\phi = f_{\text{spa}}\theta_z z_{\text{fri}}$. For the dual interference fringe, a centroid offset difference $\Delta z_{\text{fri}}$ will cause a differential phase offset, $\Delta\phi = \bar{f}_{\text{spa}}\theta_z \Delta z_{\text{fri}}$, where $\bar{f}_{\text{spa}}$ is the averaged spatial frequency. This phase shift could not be suppressed by the proposed switching methods. Considering the uncertainty of the sign of $v_{z0}$, and the installation uncertainty of the imaging system, the estimated differential phase is (0.00±0.16) rad. Detailed illustration is in the Supplementary Materials.

Imaging angle offset in the *x-y* plane will also induce differential phase error. The initial velocities of the dual atom clouds in the *x-y* plane are not exactly the same due to the imperfections of the experimental parameters. It will cause a x-direction difference $\Delta x$ of the dual-species interference fringes center. If the imaging system has a rotation angle $\theta_{\text{rot}}$ in the *x-y* plane, it will cause a differential phase offset $\Delta\phi = \bar{f}_{\text{spa}}\theta_{\text{rot}}\Delta x$. $\theta_{\text{rot}}$ is measured through the 2D interference fringe and is found to be $\theta_{\text{rot}} = (-0.034 \pm 0.011)$ rad. The value of $\Delta x$ is measured through a method that rotates the image angle of the 2D fringe image in x-y plane, and is found to be −0.175 mm. Detailed illustration is in the Supplementary Materials. The induced differential phase is calculated to be (0.009±0.003) rad. Taking account the two imaging misalignment effects, the induced differential phase is (0.01±0.16) rad.

**(7) Other Effects**

The gravity gradient induced differential phase is $\Delta\phi = k_{\text{eff}}T_{zz}\Delta z_{\text{int}}T^2$, where $T_{zz}$ is the $zz$ component of the gravity gradient tensor, which has a value of $(-2.582 \pm 0.017) \times 10^{-6}$ s$^{-2}$ according to the CSS's altitude. $\Delta z_{\text{int}}$ is the centroid difference of the dual atom interference loops, which is the same as introduced in the centrifugal force effect. This effect induces a differential phase error of $(0.0\pm1.5)\times10^{-4}$ rad.

The CSSAI uses sidebands generated by phase modulation as the Raman laser pairs. This can induce a phase shift due to the multiple-sideband effect (*57*). This phase is proportional to the value of gravity. Under microgravity condition in the CSS, this phase is suppressed by at least 3 orders compared to the case on the ground. Under actual experiment parameters, the induced phase for $^{85}$Rb and $^{87}$Rb are $\pm2.0\times10^{-5}$ rad and $\pm1.0\times10^{-5}$ rad. The induced differential phase is $(0.0\pm2.2)\times10^{-5}$ rad.

The wavefront aberration of the Raman laser will cause distortion of the shearing fringe, and induce phase shift. Because the cold atom clouds overlap and have similar expansion rates, the induced phases are common for some extent for the dual-species interference fringes (*58*). The reflecting mirror of the Raman laser has an aberration better than λ/10. To estimate the induced differential phase, we numerically generate 10 random wavefronts with λ/10 peak-to-valley distortion for the mirror and calculate the induced differential phase. The result is $(0.0\pm1.2)\times10^{-2}$ rad.

The uncertainty of the Raman laser's frequency and the interference time sequence will induce phase errors, which are proportional to the value of gravity. These phase errors are suppressed greatly due to the microgravity condition in the CSS. The laser frequency has an accuracy better than 1 MHz. The interference time sequence is referenced to a temperature-controlled crystal oscillator (TCXO), which has a frequency accuracy better than 1 ppm. The induced differential phase errors from these two effects are less than 1 μrad and can therefore be neglected.

**Acknowledgments**

    **Funding:**

This work was supported by

Space Application System of China Manned Space Program (second batch of the Scientific Experiment Project 04-3-02)

Space Application System of China Manned Space Program grant JC2-0576

Quantum Science and Technology-National Science and Technology Major Project grant 2021ZD0300603

Quantum Science and Technology-National Science and Technology Major Project grant 2021ZD0300604

Hubei Provincial Science and Technology Major Project grant ZDZX2022000001

Defense Industrial Technology Development Program grant JCKY2022130C012

National Natural Science Foundation of China grant12174403

National Natural Science Foundation of China grant W2412045

National Natural Science Foundation of China grant U25D8014

Natural Science Foundation of Hubei Province grant 2022CFA096

Strategic Priority Research Program of the Chinese Academy of Sciences grant XDA0520501

Strategic Priority Research Program of the Chinese Academy of Sciences grant XDB1070103

**Author contributions:**

Conceptualization: XC, JW, MSZ

Methodology: XC, DFZ, JTL

Investigation: XC, DFZ, JTL, WZW, WHX, JYW

Visualization: XC, DFZ

Funding acquisition: MSZ, XC, JW

Project administration: MSZ, JW, HEZ

Supervision: XC, JW, MSZ

Writing – original draft: XC, DFZ

Writing – review & editing: XC, JW, MSZ

Experimental support (optical/electronic systems, software): XL, YBW, DFG, JQZ, BT, LZ, RBL, HYS, QFC, LQ

HMLR support (microgravity preparation, command uplink): MZA, ZFL, SQW, XXG, YT, XHY

**Competing interests:** Authors declare that they have no competing interests.

**Data and materials availability:** All data are available in the main text or the supplementary materials.